\documentclass[12pt]{article}
\usepackage{epsfig}
\usepackage{latexsym,amssymb}

\def\be{\begin{equation}}
\def\beq{\begin{equation}}
\def\eeq{\end{equation}}
\newcommand{\en}{\end{equation}}
\def\ba{\begin{eqnarray}}
\def\bea{\begin{eqnarray}}
\def\ea{\end{eqnarray}}
\def\eea{\end{eqnarray}}
\newcommand{\eqa}{\begin{eqnarray}}
\newcommand{\ena}{\end{eqnarray}}

\def\Ac{{\cal{A}}}
\def\Mc{{\cal{M}}}
\def\ch{{\rm ch}}

\begin{document}
\begin{titlepage}
\vskip0.5cm
\begin{flushright}
DFTT 31/01\\
\end{flushright}
\vskip0.5cm
\begin{center}
{\Large\bf
Two-dimensional gauge theories of the symmetric group $S_n$ in the large-$n$
limit.
}
\end{center}
\vskip 1.cm
\centerline{
A. D'Adda$^a$ and P. Provero$^{b,a}$}
\vskip0.6cm
\centerline{\sl  $^a$ Istituto Nazionale di Fisica Nucleare, Sezione di Torino
and}
\centerline{\sl Dipartimento di Fisica Teorica dell'Universit\`a di Torino }
\centerline{\sl via P.Giuria 1, I-10125 Torino, Italy}
\vskip 0.2cm
\centerline{\sl $^{b}$ Dipartimento di Scienze e Tecnologie Avanzate}
\centerline{\sl Universit\`a del Piemonte Orientale}
\centerline{\sl I-15100 Alessandria, Italy
\footnote{e--mail: {\tt dadda, provero@to.infn.it}}}
\vskip 0.6cm
\begin{abstract}
We study the two-dimensional gauge theory of the symmetric group $S_n$
describing the
statistics of branched $n$-coverings of Riemann surfaces. We consider the
theory defined on the disk and on the sphere in the large-$n$ limit.
A non trivial phase structure emerges, with various phases corresponding to
different connectivity properties of the covering surface.
We show that any gauge theory on a two-dimensional surface of genus
zero is equivalent to a random walk on the gauge group manifold: in
the case of $S_n$, one of the phase transitions we find can be
interpreted as a cutoff phenomenon in the corresponding random
walk. A connection with the theory of phase transitions in random
graphs is also pointed out.
Finally we discuss how our results may be related to the known phase
transitions in Yang-Mills theory. We discover that a cutoff
transition occurs also in two dimensional Yang-Mills theory on a sphere,  
in a large $N$ limit where the coupling constant is scaled with $N$ with an 
extra $\log N$ compared to the standard 't Hooft scaling.
\end{abstract}
\end{titlepage}
\setcounter{footnote}{0}
\def\thefootnote{\arabic{footnote}}
\section{Introduction}
There are several reasons for studying gauge theories of the
symmetric group $S_n$ in the large $n$ limit. The first one
is of course that the
problem is interesting in itself,
as a simple but non trivial theory with non abelian gauge invariance.
A second reason of interest is that gauge theories of $S_n$ on a Riemann surface
describe the statistics of the $n$-coverings of the surface, namely they address
and solve the problem of counting in how many distinct ways the surface can be
covered $n$ times, without allowing folds but allowing  branch points
(see ~\cite{Kostov:1997bs,Kostov:1998bn,Billo:2001wi}
and references therein). The distinct coverings of a
two-dimensional surface are on the other hand the string configurations of a
two-dimensional string theory, so that $S_n$ gauge theories count the number of
string configurations where the world sheet wraps $n$ times around the target
space.
\par
Another reason of interest is that gauge theories of $S_n$ are
closely related to Yang-Mills theory in two dimensions, and to other
gauge models, like the one introduced by Kostov, Staudacher and
Wynter (in brief KSW) in ~\cite{Kostov:1997bs,Kostov:1998bn}.
Both YM2 ~\cite{Gross:1993hu,Gross:1993yt} and the KSW model with $U(N)$ gauge
group can in fact be
interpreted in the large $N$ limit in
terms of coverings of the two dimensional target space.
\par
In the present paper we consider the gauge theory of the symmetric group $S_n$
in its own right, in the limit where $n$, namely the world sheet
area, becomes large. The relation of this limit with the large $N$ limit of
$U(N)$ gauge theories will be discussed in Section 5.
\par
Three main results have been obtained in this paper.

The first is
the discovery of a non trivial phase  structure in the large-$n$
limit of the $S_n$ gauge theory on a sphere, with a phase transition at a
critical value of the "area" of the target surface \footnote{The exact meaning
of ``area'' in this context will be given in the next section}.
At first sight this is reminiscent of the Douglas-Kazakov
~\cite{Douglas:1993ii} phase transition for two dimensional Yang-Mills
theories, but it turns out to be a
different phenomenon, as shown in Section 5.

The second result consists in the proof of the equivalence
between gauge theories in two dimensions and random walks on group
manifolds, which, in the case of $S_n$, allows us to map our results, in
particular the aforementioned phase transition, into statements about random
walks on $S_n$. The phase transition found in the $S_n$ gauge theory
precisely corresponds to the cutoff phenomenon in random walks on
$S_n$ discovered in  ~\cite{Diaconis:1981}.

Finally we found that the same cutoff phenomenon occurs also in 2D Yang-Mills
theory with a rescaling of the coupling constant with $N$ that differ from the
standard 't Hooft rescaling by a factor $\log N$.
\par
The paper is organized as follows: in Sec. 2 we review the
correspondence between $S_n$ gauge theories and branched coverings,
defining the models we are going to study. In Sec. 3 we study the
partition function on the sphere by a saddle point analysis of the sum
over the irreducible representations of $S_n$. This allows us to identify two
lines of large-$n$ phase transitions in the phase diagram of the model. In
Sec. 4 we show the equivalence between two-dimensional gauge theories
and random walks on group manifolds. This equivalence allows us to
reinterpret the phase transition as a cutoff phenomenon in the
corresponding random walk. The approach through the
equivalent random walk is particularly suited to study the partition
function on a disk with free boundary conditions, where we find a
similar, if less rich, phase diagram. A mapping into the theory of random
graphs allows us to determine that the order parameter in the case of
the disk is the connectivity of the world sheet, and to draw some conclusions
also on the connectivity of the world sheet in the case of the sphere. In Sec. 5
we establish the relation between $S_n$ gauge theory in the large $n$ limit and
$U(N)$ gauge theories (Yang-Mills theory, chiral Yang-Mills theory and KSW
model) in the large $N$ limit and we prove the existence of a cutoff phenomenon
also for 2D Yang-Mills. Section 6 is devoted to some concluding remarks and
possible developments. Some technical details are discussed in the appendices.
\section{The model}
The statistics of the $n$-coverings of a  Riemann surface $\Mc_G$ of genus
$G$ is given by the partition function of a $S_n$ lattice gauge theory,
defined on a cell decomposition of  $\Mc_G$. This can be seen by the
following argument (more details can be found in Ref.\cite{Billo:2001wi}).
To construct a branched
$n$-covering, consider $n$ copies of each site of the cell
decomposition:
these will be the
sites of the covering surface. For each link of the target
surface, joining the sites $p_1$ and $p_2$, join each of the
$n$ copies of $p_1$ to one of the $n$ copies of $p_2$; repeat for all
the links of the target surface to define a discretized covering.
Each possible covering is defined by a choice of the copies to
be glued for each link of the target surface, that is by assigning an
element of the symmetric group $S_n$ to each link.

In general, such a covering will have branch points: consider a closed
path on the target surface, and lift it to the covering surface,
by starting on one of the $n$ sheets of the covering and changing
sheet according to the element of $S_n$
associated to the links defining the path.
The lifted path is not in general closed,
that is the covering is branched. Branch points are located on
the plaquettes of the cell decomposition, and can be classified
according to the conjugacy class of the element of $S_n$ given by
the ordered product of the elements associated to the links of the
plaquette.

For example, let $n=3$ and consider a plaquette bordered by three
links, to which the following permutations are associated
\begin{eqnarray}
P_1&=&(12)(3)\\
P_2&=&(13)(2)\\
P_3&=&(12)(3)
\end{eqnarray}
then the permutation associated to the plaquette is
\begin{equation}
P_1 P_2 P_3 = (1) (23)
\end{equation}
so that a quadratic branch point is associated to the plaquette.

The type of branch point on each plaquette is determined by the
conjugacy class of the product of the $S_n$ elements around the plaquette,
and is therefore invariant under a local $S_n$ gauge transformation
defined according to the usual rules of lattice gauge
theory. Therefore {\em a theory of $n$-coverings, in which the Boltzmann
weight depends only on the type of branch points that are present on each
plaquette, is a lattice gauge theory defined on the target
surface, with gauge group $S_n$.} The phase structure of such theories
in the large $n$ limit is the object of our study.

Let us consider first the case of unbranched coverings.
The Boltzmann weight associated to the plaquette $s$
is simply $\delta(P_s)$: the product of gauge variables around the
plaquette is constrained to be the identity\footnote{In our notation
$\delta(P)$ is one
if $P$ is the identity in $S_n$ and zero otherwise}of
$S_n$:
\be
w(P_s) = \delta(P_s) = \frac{1}{n!} \sum_r d_r \ch_r(P_s)
\label{delta1}
\eeq
In the l.h.s. of
(\ref{delta1}) the delta function is expressed as an expansion
in the characters of $S_n$, with $r$ labeling the representations of $S_n$,
$d_r$ the dimension of the representation $r$ and $\ch_r(P)$ the
character of $P$ in the representation $r$.
The partition function of this model on $\Mc_G$ is simply given
by~\cite{Billo:2001wi}:
\be
Z_{n,G}=\sum_r \left(\frac{d_r}{n!}\right)^{2-2G}
\label{nobranch}
\eeq
The partition function (\ref{nobranch})  depends only on the genus
$G$ of the surface, namely the underlying theory is a topological theory.

When branch points are allowed, the topological character of the
theory is lost, and the partition function develops a dependence on
another parameter,
which we shall call "area" and denote by ${\cal A}$, but which is not
necessarily identified with the area of $\Mc_G$. All we require is that $\Ac$ is
additive, namely that if we sew two surfaces (for instance two plaquettes)
the total "area" is the sum of the "areas" of the constituents.
In fact, by mimicking (generalized) Yang-Mills
theories, one can replace ~\cite{Billo:2001wi} the Boltzmann weight
(\ref{delta1}) of the topological theory  with a Boltzmann weight that
allows branch points:
\be
w(P_s,\Ac_s) =  \frac{1}{n!} \sum_r d_r \ch_r(P_s)  e^{{\Ac_s g_r}}
\label{plaquette}
\eeq
where $g_r$ are arbitrary coefficients and $\Ac_s$ is the area of the
plaquette $s$. The crucial property of this Boltzmann weight is that,
if $s_1$ and $s_2$ are two adjoining plaquettes and $Q$ the
permutation associated with their common link, we have:
\be
\sum_Q w(Q P_1,\Ac_{s_1}) w( P_2 Q^{-1},\Ac_{s_2}) =
w(P_1 P_2,\Ac_{s_1}+ \Ac_{s_2})
\label{cruc}
\eeq
That is by summing over $Q$ we obtain the same  Boltzmann weight that
we would have if the link corresponding to the variable $Q$ had been
suppressed  in the original lattice.

The partition function on a
Riemann surface of genus $G$ can be easily calculated from
(\ref{plaquette}) by using the orthogonality properties of the
characters:
\be
Z_{n,G}({\cal A}) = \sum_r \left(\frac{d_r}{n!}\right)^{2-2G} e^{{\cal A} g_r}
\label{zetag}
\eeq
As discussed in ~\cite{Billo:2001wi} the exponential factor in the partition
function (\ref{zetag}) can be thought of as due to a dense
distribution of branch points, in which a branch point associated to a
permutation $Q$
appears
with a probability density $g_Q$, which is related to the
coefficients $g_r$ by:
\be
g_r = \sum_{Q \neq {\bf 1}} g_{Q} \frac{\ch_r(Q)}{d_r}
\label{cierre}
\eeq
It is clear from (\ref{cierre}) that $g_Q$ is a class function,
namely it depends only on the conjugacy class of $Q$.
In the following sections we shall consider only the case in which
$g_Q=0$ for any $Q$, except the ones consisting in a single exchange.
In this case the partition function takes the form:
\be
Z_{n,G}({\cal A}) = \sum_r \left(\frac{d_r}{n!}\right)^{2-2G} e^{{\cal A}
\frac{\ch_r(\mathbf{2})}{d_r}}
\label{zetag2}
\eeq
where by $\mathbf{2}$ we denote a transposition, namely a permutation with one
cycle of length $2$ and $n-2$ cycles of length $1$. In (\ref{zetag2}) the area
has been redefined to absorb the factor $g_{ \mathbf{2}}$.
The quantity $\frac{\ch_r(\mathbf{2})}{d_r}$ at the exponent is related to the
quadratic Casimir $C_2(r)$ of a $U(N)$ representation whose Young diagram 
coincides with the one associated to the representation $r$ of $S_n$:
\be
C_2(r)= n(n-1) \frac{\ch_r(\mathbf{2})}{d_r} + nN
\label{casi}
\eeq
 So the partition function
(\ref{zetag2}) is the $S_n$ analogue of two dimensional Yang-Mills partition
function, while (\ref{zetag}) would correspond to a generalized Yang-Mills
theory.

A different way of introducing an additive parameter in the theory is to take
the "area" proportional to the number of branch points. This amounts to
expanding the partition function (\ref{zetag}) or (\ref{zetag2}) in power series
of $\Ac$ and to taking as partition function the coefficient of $\Ac^p/p!$;
in the case of only quadratic branch points the resulting partition
function is:
\be
Z_{n,G,p} = \sum_r \left(\frac{d_r}{n!}\right)^{2-2G}
\left(\frac{\ch_r(\mathbf{2})}{d_r} \right)^p
\label{zetapi}
\eeq
where, as already mentioned,  $p$ is the number of quadratic branch
points, and is identified with the "area"
\footnote{Notice that
one can only have an even number of quadratic branch points on a
closed Riemann surface, so that the partition function (\ref{zetapi}) vanish
for odd $p$.}.
In fact the partition function
(\ref{zetapi}) can be obtained by starting from a lattice consisting of $p$
plaquettes, each by definition of unit area and endowed with just one quadratic
branch point.  The Boltzmann weight of such plaquettes is:
\be
w(P_s)= \frac{1}{n!} \sum_r \ch_r(\mathbf{2})  \ch_r(P_s),
\label{plaquette2}
\eeq
If the $p$ plaquettes are joined together to form a closed surface of genus $G$,
then by using the orthogonality properties of the characters one reproduces the
partition function (\ref{zetapi}).
A  more general model can be obtained by assigning to each plaquette
a probability $x$ to have a single quadratic branch point and a probability
$1-x$ not to have any branch point at all. This amounts to consider a model with
$p$ plaquettes of Boltzmann weight
\be
w(P_s)= \frac{1}{n!} \sum_r \left( (1-x)d_r + x \ch_r(\mathbf{2})\right)
 \ch_r(P_s),
\label{plaquette3}
\eeq
which leads to the following partition function:
\be
Z_{n,G,p}(x) = \sum_r \left(\frac{d_r}{n!}\right)^{2-2G}
\left( (1-x) + x \frac{\ch_r(\mathbf{2})}{d_r} \right)^p
\label{zetapix}
\eeq
The partition function (\ref{zetapix}) includes both (\ref{zetapi}) and
(\ref{zetag2}) as particular cases. In fact for $x=1$ the partition function
$Z_{n,G,p}(x)$ coincides with $Z_{n,G,p}$ and
\be
Z_{n,G,p}(x)\stackrel {x\to 0}{\longrightarrow}
e^{-{\cal A}} Z_{n,G}({\cal A})
\label{xzero}
\eeq
with ${\cal A}= x p$.
In the following Sections we shall study the partition functions
(\ref{zetag2}), (\ref{zetapi}) and (\ref{zetapix}) in the large $n$ limit and
find that all of them show,  on a sphere,
a non trivial phase structure in the large $n$ limit:
namely they display a phase transition at a critical value of the area $p$
($\Ac$ for (\ref{zetag2})).
\par
A heuristic argument for the existence of such transition goes as follows:
consider a disk of area $p$, with holonomy at the border given by a permutation
$Q$. The corresponding partition function (consider for simplicity the case
$x=1$) is then
\be
Z_{n,{\rm disk},p}(Q) =\frac{1}{n!} \sum_r d_r \ch_r(Q)
\left(\frac{\ch_r(\mathbf{2})}{d_r} \right)^p
\label{zetadisk}
\eeq
Clearly for $p<n$ a permutation $Q$ consisting of a number of exchanges larger
than $p$ cannot be constructed out of $p$ quadratic branch points, and
$Z_{N,{\rm disk},p}(Q) $ is then necessarily zero for such $Q$.
Instead if  $p \gg n$ it is conceivable (and it will be proved in the following
sections) that all permutations have the same probability to appear on the
border and then $Z_{n,{\rm disk},p}(Q) $ is independent of $Q$ and constant in
$p$.
As a result we expect in the large $n$ limit a phase transition at a critical
value of the area, provided $p$ is rescaled with $n$ in a suitable way. We will
show that a non-trivial phase structure indeed emerges when $p$ scales
with $n \log n$ .
That is, if we put
$ p= \alpha n \log n$ a phase transition occurs in the large $n$ limit at a
critical value of $\alpha$.
As a sphere is a disk with the holonomy $Q=1$, the same phase transition appears
on the sphere as the critical point beyond which the partition function becomes
constant in $\alpha$.
\par
In the random walk approach that we will treat in Sec. 4 the same
phase transition  can be interpreted as a cutoff phenomenon, namely as
the
existence of a critical value
of the number of steps after which
the walker has the same probability to be in any point of the lattice.
The more general model (\ref{zetapix}) has, in the large $n$ limit, a phase
diagram in the $(\alpha,x)$ plane, with three different phases whose
features we shall also study in the following sections.

\section{Large $n$ limit - The variational method}
\subsection{Representations of $S_n $ in the large $n$ limit}

The large $n$ limit of the symmetric group $S_n$ is quite different
from, say, the large $N$ limit of U$(N)$. The difference
consists mainly in the fact that the
irreducible representations of SU$(N)$ are labeled by Young diagrams made by at
most $N-1$ rows and an arbitrary number of columns.  Therefore in the large
$N$ limit its rows and columns can simultaneously
scale like $N$. For instance in two
dimensional Yang Mills theory on a sphere or a  \cite{Gross:1994ub}
the saddle point at large $N$ corresponds to a representation whose Young
diagrams has a number of boxes of order $N^2$.
Instead, the irreducible representations of $S_n$ are in one to one
correspondence with the Young diagrams made of exactly $n$ boxes. Namely the
area of the Young diagrams, rather than the length of its rows and columns,
scales like $n$.

Let us first establish some notations. We shall label the lengths of the
rows of the Young diagram by the positive
integers $r_1 \geq r_2 \geq  \ldots \geq r_{s_1} $ and the lengths of the
columns by $s_1 \geq s_2 \geq \ldots \geq s_{r_1} $, with the constraint that
the total number of boxes is equal to $n$:
\be
\sum_{i=1}^{s_1} r_i = \sum_{j=1}^{r_1} s_j = n
\label{boxesn}
\eeq
In order to evaluate the partition functions introduced in the previous section,
in particular the ones given in (\ref{zetag2}),(\ref{zetapi}) and
(\ref{zetapix}), all we need is the explicit expression of the dimension of the
representation $d_r$ and of $\frac{\ch_r(\mathbf{2})}{d_r}$. These are well
known quantities in the theory of the symmetric group, and are given respectively
by:
\be
d_r = \frac{n!}{\prod_{i \leq s_j~,~j \leq r_i}(r_i+s_j-i-j+1)}
\label{dim}
\eeq
and
\be
\frac{\ch_r(\mathbf{2})}{d_r}= \frac{1}{n(n-1)} \left( \sum_i r_i^2 - \sum_j
s_j^2 \right) =  \frac{1}{n(n-1)} \left[ \sum_i \left( r_i^2 - 2 i r_i  \right)
+ n \right]
\label{cas}
\eeq
The l.h.s. of (\ref{cas}) coincides, up to a factor, with the quadratic
Casimir for the representation of a unitary group associated to the same Young
diagram.

As we are interested in the evaluation of these quantities in the large $n$
limit, we have first of all to characterize a general Young diagram
consisting of $n$ boxes in the large $n$ limit.
The most natural ansatz, in order to have a diagram of area $n$, would be to
assume that the columns $s_j$ and the rows  $r_i$ scale with $n$
respectively as $n^{\alpha}$ and $n^{1-\alpha}$ with $0 \leq \alpha \leq 1$.
However this is far from being the most general case, as different parts of the
diagram may scale with different powers $\alpha$.
To be completely general let us introduce in place of the discrete variables $i$
(resp. $j$) labeling the rows (resp. columns) the variables, continuous in the
large $n$ limit, defined by
\be
\xi = \frac{\log i}{\log n},~~~~~~~~\eta = \frac{\log j}{\log n}
\label{xsieta}
\eeq
In this way a point $(\xi_0,\eta_0)$ in the $(\xi,\eta)$ plane represents a
portion of the diagram whose rows and columns scale as $n^{\xi_0}$ and
$n^{\eta_0}$ respectively.
The Young diagram is represented by the functions
\be
\varphi (\xi) = \frac{\log r_i}{\log n},~~~~~~~~~\psi(\eta) \equiv
\varphi^{-1}(\eta) = \frac{\log s_j}{\log n}
\label{phipsi}
\eeq
and the sum over $i$ is replaced by an integral in $d \xi$:
\be
\sum_i \longrightarrow  \int d \xi ~\log n ~e^{\xi ~\log n}
\label{sumi}
\eeq
Then the constraint (\ref{boxesn}) becomes:
\be
I \equiv \int_0^{\psi(0)} d\xi ~\log n ~ e^{\log n~ \left(\xi +
\varphi(\xi) \right)} = n
\label{constr}
\eeq

Eq. (\ref{constr}) poses some restrictions on the function $\varphi(\xi)$,
namely:
\begin{description}
\item[i]  $\xi + \varphi(\xi) \leq 1$ for all $\xi$'s.
\item[ii] $\xi + \varphi(\xi) = 1 $ for at least one value of $\xi$.
\item[iii] $\xi + \varphi(\xi) = 1 $ at most in a discrete set of points.
In fact if $\xi + \varphi(\xi)=1$ in a whole interval $ a \leq \xi \leq
b$, then $I > (b-a) n \log n$.
\end{description}

\begin{figure}
\begin{center}
\epsfig{file=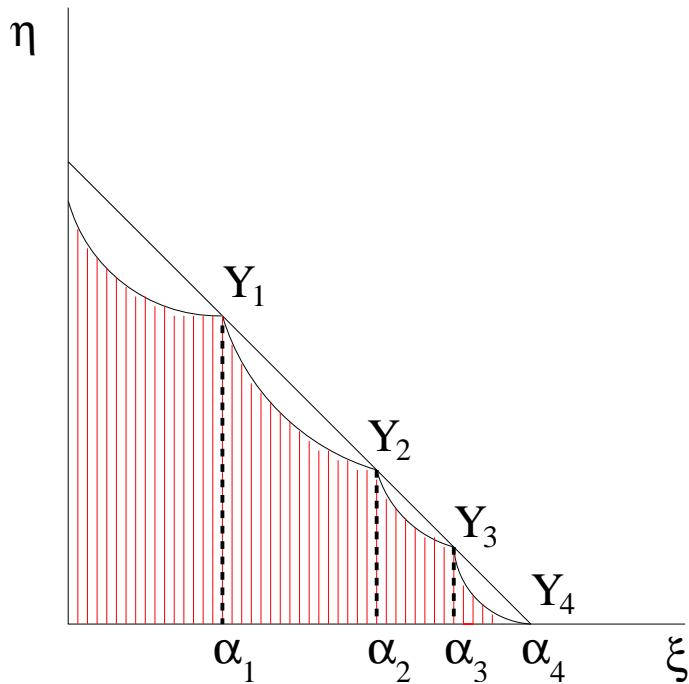,height=9.cm}
\caption{\sl
A Young diagram in the $(\xi,\eta)$ plane. Only the points touching
the line $\xi + \eta =1$ contribute in the leading order in the
large-$n$ limit.
}
\label{Figure:young}
\end{center}
\end{figure}

This means that in the large $n$ limit the contributions of order $n$ to the
l.h.s. of (\ref{constr}) come from the neighborhoods of the discrete set of
points $\alpha_t$ where $\alpha_t + \varphi(\alpha_t)=1$.
Then we can write, in the large $n$ limit:
\be
\log n~  e^{\log n (\xi + \varphi(\xi))} = n \sum_t z_t ~\delta (\xi -
\alpha_t) + o(n)
\label{delta}
\eeq
where the positive quantity $z_t$ represents the contribution to $I$ coming from
the neighborhood of $\alpha_t$, and is constrained by:
\be
\sum_t z_t = 1
\label{constr2}
\eeq
A Young diagram in the $(\xi,\eta)$ plane is represented in
Fig. \ref{Figure:young}. The
discrete set of points where the diagram touches the line $\xi + \eta =1$ are
the ones that give the $\delta$-functions in eq. (\ref{delta}), and they are the
only ones we need to consider if we restrict ourselves to the leading order in
the large $n$ limit.
Each of these points represents a portion of the diagram, which we shall denote
by $Y_t$, with area $n z_t $, and with rows and columns scaling respectively as
$n^{\alpha_t}$ and $n^{1-\alpha_t}$.
We are interested in calculating the large $n$ behavior of $d_r$ and
$\frac{\ch_r(\mathbf{2})}{d_r}$. This could be done by expressing these
quantities in terms of the functions $\varphi(\xi)$ and $\psi(\eta)$, namely by
using the logarithmically rescaled variables $\xi$ and $\eta$.
However we prefer to calculate separately the contributions
of each portion $Y_t$ by rescaling the discrete variables $i$ and $j$ with
the appropriate power of $n$, thus effectively  "blowing up" each subdiagram 
$Y_t$, which in Fig. \ref{Figure:young} 
is represented by a single point. In this way we are able
to calculate $d_r$ and $\frac{\ch_r(\mathbf{2})}{d_r}$ up to the next to leading
order, which differs from the leading one only by a $\log n$ factor and which
depends not just from the area $z_t$ of $Y_t$ but also from its "shape".
Although not relevant for obtaining the results described in the present paper
this next-to-leading order will be important for any further analysis of the
model, as discussed in the conclusions.
The details of the calculation are given in Appendix A, here we give just the
results which are useful for the present discussion. 

The large $n$ behavior of
$d_r$ is given in terms of the areas $z_t$ of the subdiagrams $Y_t$ by the
formula:
\be
\log d_r = n\log n \left\{ 1/2  z_{1/2} + \sum_{t:\alpha_t<1/2} \alpha_t
z_t + \sum_{t:\alpha_t>1/2} (1-\alpha_t) z_t \right\}+o(n\log n)
\label{dir}
\eeq
where $z_{1/2}$ is short for $z_t$ with $\alpha_t=1/2$.
As for $\frac{\ch_r(\mathbf{2})}{d_r}$ its large $n$ behavior is dominated by
the subdiagrams (which we shall denote by $Y_0$ and $Y_1$) corresponding to
a scaling power $\alpha=0$ and $\alpha=1$ respectively. All other contributions,
coming from subdiagrams $Y_t$ with $\alpha_t$ different from $0$ or $1$,
are depressed by a factor $n^{-\alpha_t}$ if $\alpha_t \leq 1/2$ or
$n^{\alpha_t - 1}$  if $\alpha_t \geq 1/2$, and can be neglected in the large
$n$ limit.
$Y_0$ (resp. $Y_1$) consists of a finite  number of columns (resp. rows)
of lengths $r_i$ (resp. $s_j$) proportional to $n$, namely:
\be
r_i = n f_i~~~~~~ s_j = n g_j
\label{fegd}
\eeq
where $f_i$ and $g_j$ are finite in the large $n$ limit.
Clearly the areas $z_0$ and
$z_1$  of $Y_0$ and $Y_1$ are given in terms of $f_i$  and $g_j$ by:
\be
z_0 = \sum_i f_i,~~~~~~z_1 = \sum_j g_j
\label{zfg}
\eeq
The large $n$ limit of the leading term of $\frac{\ch_r(\mathbf{2})}{d_r}$
can now be easily deduced from (\ref{cas}) and (\ref{fegd}) and it
is the given by:
\be
\frac{\ch_r(\mathbf{2})}{d_r} = \sum_i f_i^2 - \sum_j g_j^2 +o(1)
\label{ln}
\eeq
Notice that $z_0$ and $z_1$ do not appear at the r.h.s. of (\ref{dir}) because
their coefficients vanish. In conclusion, while the leading term of
$\frac{\ch_r(\mathbf{2})}{d_r}$ depends only on $Y_0$ and $Y_1$ the leading term
of $\log d_r$ depends only on the $Y_t$'s with $\alpha_t$ different from $0$ and
$1$.

In order to find the representation $r$ that gives the leading contribution in
the large $n$ limit to the partition functions described in the previous
section, we can then proceed in the following way: first find separately the
extrema of  $\frac{\ch_r(\mathbf{2})}{d_r}$ and $\log d_r$ at fixed  $\hat{z}$
defined by
\be
\hat{z} \equiv z_0+z_1
\label{zetahat}
\eeq
and then find the extremum in $\hat{z}$. The extrema of
$|\frac{\ch_r(\mathbf{2})}{d_r}|$ and $\log d_r$ are a direct consequence of
(\ref{ln}) and (\ref{dir}) respectively. For $|\frac{\ch_r(\mathbf{2})}{d_r}|$
the extremum corresponds either to $f_1=\hat{z}$ or to $g_1=\hat{z}$, with all
the other coefficients $f_i$ and $g_j$ equal to zero. In other words either
$Y_0$ is a diagram consisting of a single column of length $\hat{z}$, or $Y_1$
is a diagram with a single row of length $\hat{z}$. In both cases we have
\be
|\frac{\ch_r(\mathbf{2})}{d_r}|= \hat{z}^2
\label{chdue}
\eeq
For $\log d_r$ it is clear from (\ref{dir}) that the maximum, at fixed
$\hat{z}$, occurs when the whole contribution comes from a diagram with
$\alpha=1/2$, namely from a Young subdiagram where both rows and columns scale as
$n^{1/2}$. The leading term is then:
\be
\log d_r =\frac{1}{2} n \log n (1- \hat{z})
\label{drmax}
\eeq.

\subsection{Phase transitions}

Consider first the partition function (\ref{zetapi}) on a sphere,
namely at $G=0$. This can be written as:
\be
Z_{n,G=0,p} = \frac{1}{n!^2} \sum_r e^{2 \log d_r + p \log \left[
\frac{\ch_r(\mathbf{2})}{d_r} \right]}
\label{zetapisph}
\eeq
In the large $n$ limit the exponent at the r.h.s. of
(\ref{zetapisph}) can be approximated to the leading term by using
(\ref{chdue}) and (\ref{drmax}).
Moreover, as the number of representations only
grows like the number of partitions of $n$, namely in the large $n$
limit as $e^{\sqrt{n}}$, their entropy is negligible compared to the leading
term in (\ref{zetapisph}), and the sum over all representations is given by
the contribution of  the representation for which the exponent
is maximum. As discussed in the previous Section such representation is
parametrized by $\tilde{z}$, and we are led to the problem of finding the
maximum, with respect to variations of  $\tilde{z}$, of
\be
 n \log n \left[ 1 - \tilde{z} + 2 A \log \tilde{z} \right]
\label{espo}
\eeq
where, in order to have both terms of the same order in the large $n$ limit,
we have set
\be
p = A n \log n
\label{area}
\eeq
The maximum of (\ref{espo}) is at $\tilde{z}= 2A$. This solution however is
valid only for $A<1/2$, as the value of $\tilde{z}$ is limited to the interval
$(0,1)$.
So the model has two phases: for $A < 1/2$ the Young diagram of the leading
representation consists of a single row (or column) of length $2 A n$ and a
part of area $(1-2 A) n$ whose rows and columns scale like $n^{1/2}$.
For $A>1/2$ the sum over the representations is dominated by the representation
consisting of a single row (or column) of length $n$.
\par
\begin{figure}[h]
\begin{center}
\epsfig{file=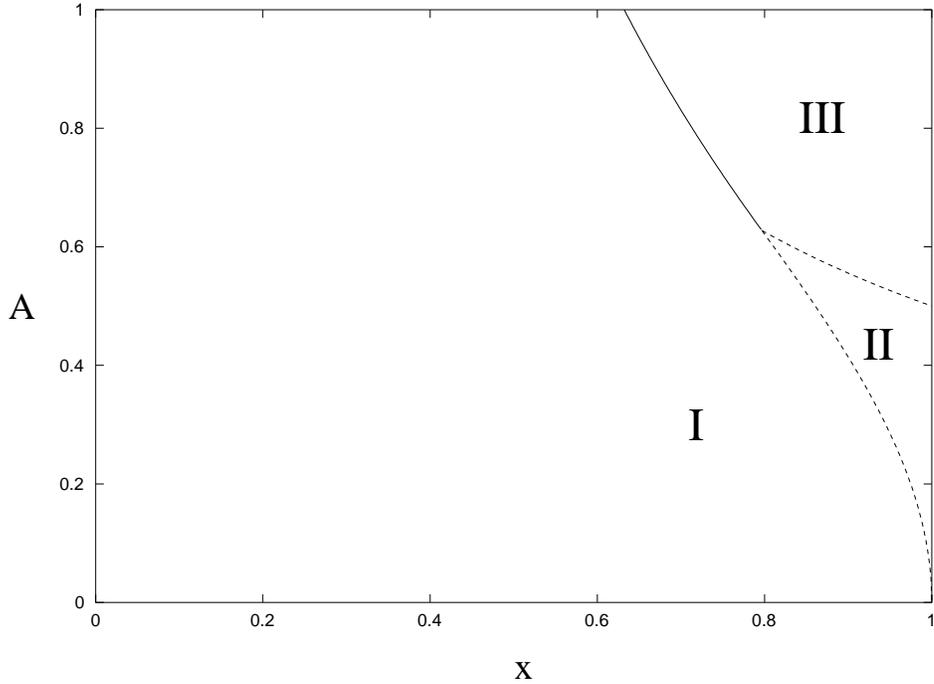,height=9.cm}
\caption{\sl
Phase diagram in the $(x,A)$ plane. The phase transitions between
phases I/III and II/III are first order, while the II/III transition
is second order.}
\label{Figure:phase}
\end{center}
\end{figure}

Let us consider now the more general model whose partition function is given in
(\ref{zetapix}). A difference with respect to the previous case is that $p$ does
not need to be even, and the symmetry with respect to the exchange of rows and
columns in the representations is broken. So for each value of $\tilde{z}$ there
are two representations, whose contribution differ for the sign of
$ \ch_r(\mathbf{2})$. It easy to check from (\ref{zetapix}) that the
contributions coming from representations with positive sign of
$ \ch_r(\mathbf{2})$ are always greater in absolute value, and give rise to the
leading term in the large $n$ limit. The leading representation is then obtained
by finding the value of $\tilde{z}$, constrained by $0 \leq \tilde{z} \leq 1$,
for which
\be
n\log n \left[ 1-\hat{z} + A \log \left( 1-x+x \hat{z}^2 \right)
\right]
\label{elibx}
\eeq
is maximum.
The phase structure is more complicated than in the previous case, and it is
represented in the $(x,A)$ plane in Fig. \ref{Figure:phase}.
The phase labeled in the figure
with I corresponds to $\hat{z}=0$, namely to a situation where the dominant
representation is entirely made of rows and columns that scale as $n^{1/2}$.
This phase did not exist in the previous case ($x=1$) except for the trivial
point $A=0$.
Phase II and III are the ones already studied at $x=1$ and correspond
respectively to $0<\tilde{z}<1$ and $\tilde{z}=1$.
The critical line that separates I from III can be easily calculated, and is
given by:
\be
A_{I,III}= -\frac{1}{\log (1-x)}
\label{crita}
\eeq
while the critical line separating phase II and III is simply:
\be
A_{II,III} = \frac{1}{2 x}
\label{crita2}
\eeq
Finally the line that separates phase I and phase II is given by
\be
A_{I,II}= \sqrt{\frac{1-x}{\varphi x}}
\label{crita3}
\eeq
where $\varphi$ can be expressed in terms of the coordinate $x_c$ of the triple
point: $\varphi=4 x_c (1-x_c)$. The triple point is at the intersection of
(\ref{crita}) and (\ref{crita2})and its coordinate $x_c$ is the solution of the
transcendental equation $\log (1-x) + 2 x =0$.
Its numerical value is $x_c=0.796812..$, from which one also obtains
$\varphi=0.647611..$.
The critical line (\ref{crita2}) is what one  expects, according to the results
of the $x=1$ model, from an effective number of branch points equal to the 
number of plaquettes $A n \log n$ times the probability $x$ for a plaquette to 
have a branch point.  
However the phase diagram discussed above shows that such naive
expectation is not fulfilled everywhere. This can be understood by calculating
the large $n$ limit of (\ref{zetapix}) in a slightly different way.
We first expand the binomial in (\ref{zetapix}) and write:
\ba
Z_{n,G=0,p}(x)&=&\sum_r \sum_{k=0}^{p} \left( \frac{d_r}{n!} \right)^2
\left( \begin{array}{c} p \\ k \end{array} \right) (1-x)^{p-k} x^k
\left(\frac{\ch_r(\mathbf{2})}{d_r}\right)^k \nonumber \\&=& \sum_{k=0}^{p}
\left( \begin{array}{c} p \\ k \end{array} \right) (1-x)^{p-k} x^k
Z_{n,G=0,k}
\label{zpx}
\ea
In the large $n$ limit we parametrize $p$ as in (\ref{area}), and $k$ as
$k = \lambda p$. The sum over $k$ is replaced by an integral over $\lambda$ that
can be evaluated using the saddle point method. In doing this the large $n$
solution for $Z_{n,G=0,k}$ must be used, keeping in mind that this consists of
two phases, one for $2 \lambda A>1$ and one for$2 \lambda A<1$. The calculation
reproduces the phase diagram of Fig. \ref{Figure:phase}. The saddle point corresponds to
$\lambda=0$ in phase I, to $\lambda=x$ in phase III and to $0<\lambda<x$ in
phase II.
The free energy in the different phases can be obtained from (\ref{elibx}) by
replacing $\tilde{z}$ with the relevant saddle point solution. So we have:
\ba
F_n(A,x) = n \log n \left[ 1 - z(A,x) + A \log [1-x+x z(A,x)^2] \right]
\label{freen}
\ea
where  $z(A,x)=0$ if the point $(A,x)$ is in I, $z(A,x)=1$ if $(A,x)$ is in III,
while for $(A,x)$ in II we have
\ba
z(A,x) = A+ \sqrt{A^2 - \frac{1-x}{x}}
\label{zetasad}
\ea
As the free energy is known exactly in the large $n$ limit in any point of the
$(A,x)$ plane, the order of the phase transitions can be explicitly calculated.
The (I,II) and the (I,III) phase transitions are of first order, while the
(II,III) phase transition is a second order phase transition with the second
derivative of $F_n(A,x)$ with respect to $A$ finite everywhere but with a
discontinuity at the critical point $A=1/2x$.
The three phases are characterized by different connectivity properties of the
world sheet. We have not been able to investigate these properties within the
variational approach, so we have to rely on the equivalence between gauge
theories of $S_n$ and random walks on one hand, and between random walks and
random graphs on the other. These will be discussed in the following sections;
in particular the connectivity of the world sheet in the different phases is
discussed in Section 4.3, as a corollary of well known results in the theory of
random graphs.

Finally let us consider the model given by the partition function
(\ref{zetag2}). As already pointed out, this coincides with the small $x$ limit
of (\ref{zetapix}) provided we put ${\cal A}= x p= x A n \log n$. For small $x$
the first order phase transition occurs at $A=-\frac{1}{\log (1-x)}$, hence
for (\ref{zetag2}) at ${\cal A}=  n \log n$.

\section{Gauge theories on a disk as random walks on the group manifold}
A two-dimensional gauge theory on a disk is equivalent to a random
walk on the gauge group manifold, the area of the disk being
identified with the number of steps and the gauge theory action with
the transition probability at each step.
\par
This result is completely general with respect to the choice of
the gauge group and of the action, as we will prove below. However it
might be useful to show first how this result emerges in the $S_n$ gauge
theory defined by Eq.(\ref{zetapi}), that is in a theory where all plaquettes
variables are forced to be equal to  transpositions.
\par
\begin{figure}[ht]
\begin{center}
\epsfig{file=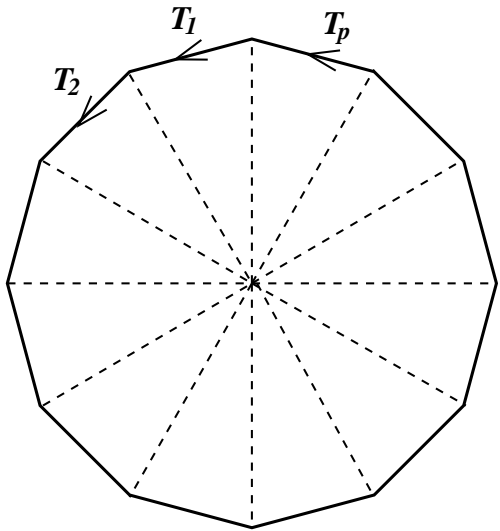,height=7.cm}
\caption{\sl A cell decomposition of the disk. The permutations on the
dashed links have been gauge-fixed to the identity, so that the ones
on the boundary are forced to be transpositions. The ordered product
of the $p$ transpositions gives the holonomy.}
\label{Figure:disk}
\end{center}
\end{figure}

Suppose we want to compute such partition function on a disk with a
fixed holonomy $Q\in S_n$, Eq. (\ref{zetadisk}). The latter shows that
the partition function depends only on the holonomy $Q$ and the area
of the disk (that is the theory is invariant for area-preserving
diffeomorphism).
It follows that we can freely choose any cell decomposition of the
disk made of $p$ plaquettes, {\em e.g.} the one shown in
Fig. \ref{Figure:disk}.
To compute the partition function means to count the ways in which we
can place permutations $P$ on all the links in such a way that
\begin{itemize}
\item
the ordered product of the links around each plaquette is a
transposition
\item
the ordered product of the links around the boundary of the disk is a
permutation in the same conjugacy class as $Q$
\end{itemize}
\par
Now we can use the gauge invariance of the theory to fix all the
radial links to contain the identical permutation. At this point the links
on the boundary are forced to contain transpositions: therefore
the partition function with holonomy $Q$ is the number
of ways in which one can write the permutation $Q$ as an ordered
product of $p$ transpositions. This in turn can be seen as a random walk
on $S_n$, in which, at each step, the permutation is multiplied by a
transposition chosen at random: the gauge theory partition function
for area $p$ and holonomy $Q$ is the probability that after $p$ steps
the walker is in $Q$.
\subsection{The correspondence for a general gauge theory}
To show that this result actually holds for all gauge groups and
choice of the action, consider now a gauge theory on a disk of area
$p$ with gauge
group $G$ and holonomy $g\in G$ on the disk boundary.
To fix the notations, we will consider a finite group $G$, but the
argument can be extended to Lie groups.
The theory is defined by a function $w(g)$ such that the
Boltzmann weight of a configuration is given by the product of
$w(g_{pl})$ over all plaquettes of the lattice, with $g_{pl}$ the
ordered product of the links around the plaquette. For the theory to
be gauge invariant, $w$ has to be a class function; moreover we will
require $w(g) \ge 0$ for all $g$ and normalize $w$ so that  $\sum_g
w(g)=1$.
\par
The partition function is \cite{Migdal:1975zg, Rusakov:1990rs}
\be
Z_p(g)=\frac{1}{|G|}\sum_r d_r \tilde{w}_r^p\chi_r(g)
\eeq
where the sum is over all irreducible representations of $G$,
$\chi_r(g)$ is the character of $g$ in the representation $r$, and the
$\tilde{w}_r$'s are the coefficients of the character expansion of the
Boltzmann weight:
\be
w(g)=\frac{1}{|G|}\sum_r d_r \tilde{w}_r \chi_r(g)
\eeq
\par
Now consider a random walk on $G$ with transition probability defined
as follows: if the walker is in $g_p\in G$ at step $p$, then its
position at step $p+1$ is obtained by left multiplying $g_p$ by an
element $g$ chosen in $G$ with a probability $t(g)$ which is a class
function, {\em i.e.} depends on the conjugacy class of $g$ only.
\par
Suppose the random walks starts in the identity of $G$, and call
$K_p(g)$ the probability that the walker is in $g$ after the $p$-th
step. Then
\be
K_{p+1}(g)=\sum_{g^\prime} t(g^\prime g^{-1}) K_p(g^\prime)
\label{recursion}
\eeq
Now assume $K_p(g)$ is a class function, with character expansion
\be
K_p(g)=\frac{1}{|G|}\sum_r d_r k^{(p)}_r \chi_r(g)
\eeq
then it follows form Eq.~(\ref{recursion}) that also $K_{p+1}$ is
a class function, and the coefficients of its character expansion are
\be
k^{(p+1)}_r=\tilde{t}_r k^{(p)}_r
\eeq
where the $\tilde{t}_r$'s are the coefficients of the character expansion of
the class function $t$:
\be
t(g)=\frac{1}{|G|}\sum_r d_r \tilde{t}_r \chi_r(g)
\eeq
Now, since $K_1(g)=t(g)$, it follows by induction that $K_p$ is indeed a class
function, and
\be
k^{(p)}_r=\tilde{t}_r^p
\eeq
so that the probability distribution after $p$ steps
of the random walk
equals the gauge
theory partition function $Z_p(g)$ provided the Boltzmann weight of
the plaquette in the latter is identified with the transition function
of the former:
\be
w(g)=t(g)
\eeq
\par
In conclusion, the partition function of a gauge theory on a disk,
of area $p$ with
a certain holonomy $g$ on the disk boundary, equals the probability
that a random walk that starts in the identity of the group will reach
the element $g$ in $p$ steps, each step consisting of left
multiplication by an element chosen with a probability distribution
coinciding with the plaquette Boltzmann weight of the gauge theory.
\subsection{Cutoff phenomenon in random walks on $S_n$}
The cutoff phenomenon in random walks was discovered in
Ref.\cite{Diaconis:1981}, where a random walk on $S_n$ was studied in
which at each step the permutation is multiplied by the identical
permutation with probability $1/n$ and by a randomly chosen
transposition otherwise. According to the argument of the previous
section, this corresponds to our model Eq.(\ref{zetapix}) with
$x=1-1/n$.
The holonomy of the gauge theory translates into constraints on the
element of $S_n$ where the random walk ends: for example the partition
function on the sphere will count the walks that return to the
identical permutation in $p$ steps.
\par
The main result of Ref.\cite{Diaconis:1981} is that if the number of
steps scales as $A n \log n$, then in the large-$n$ limit for $A>1/2$
the probability of finding the walker in any given element $Q\in S_n$
is just $1/n!$ for all $Q$: complete randomization has been achieved
and all memory of the initial position of the walker has been erased.
\par
In terms of the corresponding gauge theory, this can be translated
into a statement
about the partition function on a disk: for $A>1/2$, the partition
function with any given holonomy $Q$ stops depending  on $A$ and is simply
proportional to the number of permutations in the conjugacy class of
$Q$. This is true in particular for $Q=1$, corresponding to the
partition function on the sphere. Therefore the phase transition found
in Sec. 3 has a natural interpretation as a cutoff phenomenon in the
corresponding random walk.
\par
Strictly speaking, this applies only to the specific model $x=1-1/n$
studied in Ref.\cite{Diaconis:1981}. However we want to argue that
this is the correct interpretation of the whole line of phase
transitions at $A=\frac{1}{2x}$ found in Sec. 3.
Consider first the model with $x=1$, where at each step the
permutation is multiplied by a random transposition. The probability
distribution does not have a limit as the number of steps goes to
infinity, since for even number of steps one can only obtain an even
permutation and {\em vice versa}. Therefore the probability distribution in
$S_n$ can never become uniform. However it is natural to expect
that a sort of cutoff phenomenon occurs all the same at number of
steps $p=1/2 \ n \log n$, and precisely that for even $p$ the probability
distribution becomes uniform in the {\em alternating} group and  for
odd $p$ in its
complement.
\par
To support this conjecture, let us compute {\em e.g.} the expected number of
cycles of length 1 in the permutation obtained after $p$ steps.
The calculation is described in Appendix B, and the result is
\be
N_1(x=1,p)=1+(n-1)\left(\frac{n-3}{n-1}\right)^p
\eeq
so that for $p=A n \log n$ we have
\be
N_1(x=1)\sim \left\{ \begin{array}{cc} n^{1-2A}&{\rm for}\ A\le 1/2\\
                                      1        &{\rm for}\ A>1/2
\end{array}\right.
\label{N1}
\eeq
The result for $A>1/2$ is the one expected for a uniform probability
distribution in the alternating group or its complement.
\par
Repeating the calculation for arbitrary $x$ one finds
\be
N_1(x)\sim \left\{ \begin{array}{cc} n^{1-2xA}&{\rm for}\ A\le 1/(2x)\\
                                      1        &{\rm for}\ A>1/(2x)
\end{array}\right.
\label{N1x}
\eeq
so that the cutoff phenomenon occurs at $A=1/(2x)$, the result one
intuitively expects from the fact that a fraction $x$ of the random
walk steps are ``wasted'' in doing nothing and do not contribute to
the randomization process.
\subsection{Results from random graphs theory}
In the previous sections we have mainly considered the theory defined
on a sphere: we have found a complex phase structure with first and
second-order phase transitions. One of the transition lines can be
interpreted as a cutoff phenomenon in the corresponding random walk.
\par
In this section we consider the theory with free boundary conditions:
the permutations on the boundary of the disk are summed over like the
internal ones.
From the point of view of the random walk, this implies considering
all the possible paths irrespective of the permutation they end in
after $p$ steps.
We will exploit a correspondence between cutoff
phenomena in random walks and phase transitions in random graphs first
noted in Ref.~\cite{Pak:2001}.
Consider the random walk in $S_n$ defined by $x=1$, {\em i.e.}
at each step the permutation is multiplied by a random
transposition (but all the arguments we will give translate trivially to
the case $x<1$). From the point of view of coverings,
a step in which the transposition $(ij)$ is used corresponds to adding a
simple branch point that connects the two sheets
$i$ and $j$ of the covering surface. One can think
of the process as
the construction of a random graph on $n$ sites where at each step  a
link between two sites is added at random. After $p=A n \log n$
steps the expected number of links is equal to the number of
steps (since the number of available links is $O(n^2)$ the fact that
the same link can be added more than once can be neglected in the
large $n$ limit; see Ref.~\cite{Pak:2001}).
\par
It is a classic result in the theory of random graphs
\cite{Erdos:1959,Erdos:1973}
that if the
number of links $p$ is smaller than $1/2\  n \log n$ then the graph is
almost certainly disconnected while for $p>1/2\  n \log n$ the graph
is almost certainly connected, where ``almost certainly'' means that
the probability is one in the limit $n\to \infty$.
Therefore we conclude that the model with free boundary conditions
undergoes a phase transition at $A=1/2$ where the covering surface goes
from disconnected  to connected. For $x<1$, the same transition occurs
at $A=1/(2x)$.
\par
Notice that the free boundary conditions are crucial for this argument
to work: in the case, say, of the sphere, the corresponding random
walk is forced to go back to the initial position in $p$ steps, so
that links in the graph are not added independently and the result of
Ref.\cite{Erdos:1959,Erdos:1973} do not apply. However some consequences can be
drawn from these results also for the case of the sphere. Consider first
the model with $x=1$ and a sphere of area $A>1$.
The latter can be thought of as two disks, both
with area $A>1/2$, joined together. The partition function of the
sphere is then obtained by multiplying together the partition
functions of the two disks and by summing over the common holonomy $P$ on
the border. If the world sheets of the two disks are almost certainly
connected, the same applies to the world sheet of the resulting
sphere. Strictly speaking this proof holds only for $A>1$, however we
have shown that the leading term of order $n \log n$ of the free
energy is independent of $A$ for $A>1/2$. So unless a phase
transition occurs due to the next leading term of order $n$ (which
cannot be ruled out {\it a priori}) the whole phase with $A>1/2$ at
$x=1$ (and extending the argument to $x<1$ the whole phase III) will
be characterized by a connected world sheet. What can be said about
phase II and I? We mentioned already the result in the theory of
random graphs that for a number of links $p$ smaller than $1/2 n \log n$
the graph is almost certainly disconnected. Although this result
applies to the case of free boundary conditions, it should {\it a
fortiori} be true also for the sphere. In fact the sphere corresponds
to a random walk which is forced to end in the identical
permutation, thus favoring graphs in which less links are turned on.
So for $A<1/2$  at $x=1$, namely in phase II, and {\it a fortiori} in
phase I we expect the world sheet to be disconnected.
Another result in  random graphs \cite{Erdos:1959,Erdos:1973} states that if the
number of links $p$ grows like $\epsilon n \log n$ with $\epsilon>0$,
the size $n_c$ of the largest connected graph is $n$ in the large $n$
limit, namely $\lim_{n \rightarrow \infty} n_c/n =1$. Again, although
the result is proved for graphs corresponding to random walks with
free boundary conditions, it can be extended, at $x=1$, to a sphere of area
$\epsilon n \log n$ which can be thought of as obtained by sewing  two disks
of area $\epsilon /2 n \log n$. Both phase III and phase II are then
characterized by the presence of a connected world sheet of size $n_c\sim  n$
in the large $n$ limit, but phase III has completely connected world sheets
while phase II has not.
In phase I the number of effective branch points grows slower than
any $\epsilon n \log n$, possibly like $\alpha n$. If that is the case (a
detailed analysis of next to leading terms would be required to check this
point) then another result \cite{Erdos:1959,Erdos:1973} of random graphs could
be applied. This states that if the number of links is $\alpha n$ in the large
$n$ limit, then for $\alpha>1/2$ the largest connected part has size
$\psi ( \alpha) n$ with $\psi(1/2)=0$ and $\psi(\infty) =1$. The function
$\psi (\alpha)$ is known and can be written as an infinite series.
The point $\alpha=1/2$ is the percolation threshold, its existence may be an
indication of further phase structure within phase I.

\section{Relation with two dimensional Yang-Mills theories}
Besides being linked to the theory of random walks and random graphs,
$S_n$ gauge theory is also closely related to lattice $U(N)$ gauge theories
on a Riemann surface. This was first discovered by Gross and Taylor
~\cite{Gross:1993hu,Gross:1993yt} ,
who found that the coefficients of the large $N$ expansion of the $U(N)$
partition function could be interpreted in terms of string configurations,
namely of coverings of the Riemann surface.
In the case of  $U(N)$ Yang-Mills theory the maps from
the string world sheet to the target space have two possible orientations and
world sheets of opposite orientations can interact only through point-like
singularities. As a result the theory almost exactly factorizes into two copies
of a simpler, orientation preserving chiral theory. This chiral Yang-Mills
theory is obtained as a
truncation of the whole theory by restricting the sum over the irreducible
representations of $U(N)$ to the representations
whose Young diagram contains a  finite number of
boxes in the large $N$ limit.
The number $n$ of boxes in a
representation coincides with the  number of times the world sheet of
the corresponding string configuration covers the target space.
While in the gauge theory of $S_n$ the irreducible representations are labeled
by Young diagrams that contain exactly $n$ boxes, in chiral $U(N)$ gauge theory
the irreducible representations are labeled by Young diagrams which contain an
arbitrary number of boxes and are only restricted by the condition that the
number of rows do not exceed $N-1$.

The partition function of chiral Yang-Mills theory can then be written as a sum
over $n$, and we expect each term in the sum, being the number of $n$-coverings,
to be related to an $S_n$ gauge theory.
For chiral Yang-Mills this is  true  only on a torus, namely if the genus
of the target space is zero. It is well known in fact that for
different genuses the coefficients of the $1/N$ expansion of the
$U(N)$ partition function are not directly related to the number of
coverings, due to presence of the so called $\Omega^{-1}$ points.

A matrix model that gives, to all orders in the $1/N$ expansion, the  exact
statistic of branched coverings on a Riemann surface and whose restriction to a
fixed value of $n$ coincides with a $S_n$ gauge theory was introduced by
Kostov, Staudacher and Wynter (KSW in short) in
\cite{Kostov:1997bs,Kostov:1998bn}. This model has
still a $U(N)$ gauge invariance but realized in terms of complex rather
than unitary $N\times N$ matrices.
The partition function of the KSW model on a Riemann surface of genus $G$ is
\cite{Kostov:1997bs,Kostov:1998bn}\footnote{We have restricted the KSW model to
contain only quadratic branch points, for the general case see the original
papers}:
\be
Z_{N,G}^{(KSW)}(\tau,\mu)=\sum_n \sum_{r \in U(N),|r|=n} \left(N^n \frac{d_r}{n!}
\right)^{2-2G} \exp \left[\frac{\tau n(n-1)}{2 N} \frac{\ch_r(\mathbf{2})}{ d_r} - n
\mu \right]
\label{ksw}
\eeq
where we denote by $r$ both the Young diagrams and the corresponding
representations of either $U(N)$ of $S_{|r|}$, with $|r|$ the number of boxes
in $r$. The sum is over all Young diagrams corresponding to representations of
$U(N)$, $d_r$ and $\ch_r(\mathbf{2})$ are the same as in the previous sections.
By comparing (\ref{ksw}) with (\ref{zetag2}) we can write:
\ba
Z_{N,G}^{(KSW)}(\tau,\mu)&=& Z_{1,G}({\cal A}) N^{2-2G} e^{-\mu}+
Z_{1,G}({\cal A}) N^{2(2-2G)} e^{-2 \mu}+\ldots \nonumber \\
&+&Z_{N,G}({\cal A}) N^{N(2-2G)} e^{-N \mu}+ \tilde{Z}_{N+1,G}({\cal A})
N^{(N+1)(2-2G)} e^{-(N+1)\mu}+ \ldots
\label{grotay}
\ea
\par
The partition functions at the r.h.s. of (\ref{grotay}), for $ n \leq N$, are the
partition functions of the $S_n$ gauge theory given in (\ref{zetag2}) with
${\cal A}= \frac{\tau n (n-1)}{2 N}$. For $n>N$ the partition functions, denoted
by $\tilde{Z}$, are {\it incomplete} $S_n$ partition functions, because in that
case some irreducible representations of $S_n$, whose Young diagram  has more
than $N$ rows, do not correspond to any representation of $U(N)$.

The partition function of chiral Yang-Mills theory is similar to
(\ref{ksw}), but with the dimension $\Delta_r$ of the $U(N)$
representation replacing the factor $N^n \frac{d_r}{n!}$ at the
r.h.s.. The ratio between these two factors is the so called $\Omega_r$
term, whose presence prevents chiral Yang-Mills theory from having a
simple interpretation in terms of coverings for $G \neq 1$.

In spite of these differences two dimensional Yang-Mills theory,
chiral Yang-Mills theory and the KSW model all share some common
features in the large $N$ limit.
If the target space has the topology of a sphere ($G=0$) all these
theories exhibit a non trivial phase structure in the large $N$ limit. In the
case of two dimensional Yang-Mills theory there is a third order phase
transition, the Douglas-Kazakov phase transition ~\cite{Douglas:1993ii}. This
occurs at a critical value $\Ac = \pi^2$ of the area of the sphere, measured in
units of the coupling constant. This phase transition is well understood in terms
topologically non trivial configurations ~\cite{Caselle:1993mq,Gross:1994mr}.
The phase structure of chiral Yang-Mills theory and of the KSW model has been
studied in ~\cite{Kostov:1998bn} and in ~\cite{Crescimanno:1994eg} and it turns
out to be richer than pure Yang-Mills. In the KSW model four distinct
phases are present in the $(\tau,\mu)$ plane.
In all these cases the saddle point in the large $N$ limit corresponds to a
representation of $U(N)$ where  both row and columns
of the associated Young diagram scale as $N$, hence the total number
of boxes is of order $N^2$:
\be
n \propto N^2
\label{propto}
\eeq
This means that the saddle point at large $N$  corresponds to a
string configuration that covers the target space a number of times
$n$ proportional to $N^2$, and that the associated Young diagram has
rows and columns of order $\sqrt{n}$.
Besides the free energy is of order $N^2$, namely of order $n$. This
also means that the number of branch points $p$ if of order $n$. In
fact the free energy is  the exponent at the r.h.s. of
(\ref{ksw}) calculated at the saddle point plus a term, that does not
depend from the number of branch points, coming from the dimension of the
representation.
Let us expand the exponential in (\ref{ksw}) in power series:
\be
\exp \left[\frac{\tau n(n-1)}{2 N} \frac{\ch_r(\mathbf{2})}{ d_r} - n
\mu \right] = \sum_p \frac{1}{p!}  \left[\frac{\tau n(n-1)}{2 N} 
\frac{\ch_r(\mathbf{2})}{ d_r} - n \mu \right]^p
\label{expo}
\eeq
Each term in the sum at the r.h.s. corresponds to a configuration with $p$
plaquettes, each plaquette with either a single quadratic branch point or no
branch point at all with probabilities respectively proportional to 
$\frac{\tau n(n-1)}{2 N}$ and $n \mu$.
The former grows faster with $n$ than the latter, so in the large $n$ limit the
probability $x$ of having a branch point in each plaquette tends to 1.

Finally we remark that the sum over $p$ in (\ref{expo}) is dominated at large
$n$ by by a single value of $p$\footnote{The saddle point of $\sum_j
\frac{x^j}{j!}$ in the large $x$ limit is $j=x$, as shown by Stirling formula.}
, namely $p= \left[\frac{\tau n(n-1)}{2 N} \frac{\ch_r(\mathbf{2})}{ d_r} - 
n \mu \right]$. This implies that $p$ is also of order $n$\footnote{Remember
that the saddle point is at $N \propto \sqrt{n}$ and $\frac{\ch_r(\mathbf{2})}
{ d_r} \propto n^{-1/2}$}.
To summarize: the standard large $N$ limit of $U(N)$ gauge theories is
described in terms of string configurations (coverings), which are also
configurations of an $S_n$ gauge theory, with $n$ given by (\ref{propto}),
rows and columns scaling like $\sqrt{n}$ and
\be
p \propto n,~~~~~~~~~x=1
\label{propto2}
\eeq
That means the large $N$ limit of $U(N)$ gauge theories corresponds to the
point at the right corner of
region I in Fig. \ref{Figure:phase}, 
where the coefficient of $n \log n $ in $p$ is strictly
zero.
In order to "blow up" that point and determine weather a further phase structure
is present there one would need to evaluate the terms of order $n$ in the free
energy. This will be the task of a future work.
We just remark here that the presence of a non trivial phase structure in that
region is likely for at least two reasons: because that region is the section 
with a
constant $n$ plane of the large $N$ limit of the KSW model, which has a non
trivial phase structure, and because we know from the random graphs theory that
for $p \propto n$ there is at least one transition, the percolation transition.

It appears from the previous discussion that the phase transition that we
found in the $S_n$ theory, and that corresponds to the well known cutoff
phenomenon in random walk, does not have a counterpart among the
known phase transitions in two dimensional Yang-Mills theory.
It is quite natural to ask weather a transition of this kind exists also
for two dimensional Yang-Mills and other $U(N)$ gauge theory on a sphere.
We found the answer to be affirmative. This new type of phase transition,
the cutoff phenomenon, can be observed  provided the coupling constant is rescaled
with $N$ with an extra $\log N$ with respect to the usual 't Hooft prescription.
We give here a simple, although rigorous, argument, leaving a
detailed analysis of the transition to a future work.
Consider the partition function of YM2 on a sphere (see for instance
~\cite{Gross:1994mr}) with gauge group $U(N)$:
\be
Z_0(A,N) = e^{-\frac{A}{24}(N^2 - 1)} \sum_{n_1>n_2>...>n_N} \Delta^2
(n_1,...,n_N) e^{-\frac{A}{2 N} \sum_{i=1}^N n_i^2}
\label{qcdsph}
\eeq
where the $n_i$'s are integers, $\Delta(n_1,...,n_N)$ is the Vandermonde
determinant and $A$ the area of the sphere. In the large $N$ limit a la
't Hooft $A$ is kept fixed. We are going to allow $A$ to rescale with $N$:
$A \rightarrow A(N)$. For $A(N)$ sufficiently large we expect the sum to be
dominated in the large $N$ limit by the configuration for which the exponential
is maximum, namely $\{n_1,n_2,.....n_N\}=\{\frac{N-1}{2},\frac{N-1}{2} -1,....
-\frac{N-1}{2}\}$. This configuration corresponds to the trivial representation
of $U(N)$, and it is the exact analogue of the representation of $S_n$
consisting of a single row. Let us determine now the value of $A(N)$ for which
this configuration ceases to be a maximum by comparing its contribution to the
sum in (\ref{qcdsph}) with the contribution of a configuration where $n_1$ is
increased of one unit. We find:
\be
\frac{\Delta^2 (n_1,...,n_N) e^{-\frac{A}{2 N} \sum_{i=1}^N n_i^2}
|_{\{n_1,n_2,.....n_N\}=\{\frac{N-1}{2},\frac{N-1}{2} -1,....-\frac{N-1}{2}\}}}
{\Delta^2 (n_1,...,n_N) e^{-\frac{A}{2 N} \sum_{i=1}^N n_i^2}
|_{\{n_1,n_2,.....n_N\}=\{\frac{N-1}{2}+1,\frac{N-1}{2}
-1,....-\frac{N-1}{2}\}}}= \frac{e^{\frac{ A(N)}{2}}}{N^2}
\label{clog}
\eeq
The cutoff phenomenon occurs when the r.h.s. of (\ref{clog}) is greater than
$1$, namely for $A(N) > 4 \log N$, while for  $A(N) < 4 \log N$ we
are in presence of a new phase whose characteristics are still to be
determined.

\section{Conclusions}
Let us summarize the results obtained in the paper.
We have studied a two-dimensional gauge theory of the symmetric group
$S_n$ that describes the statistics of branched coverings on a
Riemann surface, in the large-n limit.
\begin{itemize}
\item
The theory on the sphere shows
an interesting phase diagram when the number of branch points scales
as $n \log n$, with lines of first and second-order phase transitions.
\item
All two-dimensional gauge theories on a genus-0 surface can be mapped
into random walks in the corresponding group manifold. In our case,
this allows us to interpret one of the transition lines as a cutoff
phenomenon in the corresponding random walk.
\item
The theory on a disk, with free boundary conditions, can be studied
with methods of the theory of random graphs: this allows one to
show that there is a phase transition on a disk from a disconnected to
a connected covering surface. From this one can argue, and with some limitations
prove, that the connectedness of the covering is what characterizes the different
phases also on the sphere.
\item
A cutoff phenomenon is found also in 2D Yang-Mills on a sphere, if
the area of the sphere scales with $N$ like $N \log N$
\end{itemize}

\par
The present paper can be extended in two distinct directions. On one hand
it would be desirable to understand better region I of the phase
diagram of Fig. \ref{Figure:phase}, and in particular its right corner
which corresponds to the large $N$ limit 
of $U(N)$ gauge theories. For this purpose the variational approach should be
implemented to include the contributions of the next-to-leading order  in $n$,
which is of order $n$ instead of $n \log n$. This might reveal further phase
structure, like for instance a percolation phase transition. 
Also,  phase transitions in region I should be the analogue in $S_n$ gauge 
theory of the Douglas-Kazakov phase transition in 2D Yang-Mills, and of the 
phase transitions studied in \cite{Crescimanno:1994eg} and \cite{Kostov:1998bn} 
for chiral Yang-Mills and KSW model respectively.
It can be shown that the calculation of the correlators, which are relevant
in order to determine the order parameters of the various phase transitions, 
also requires to know the free energy beyond the leading order.

The existence of  a cutoff phenomenon in two dimensional Yang-Mills, although
within the framework of a non conventional scaling with $N$ of the coupling 
constant, is also a fact whose meaning and physical implications, if any, should
be further investigated. 
In particular a full description of the phase preceding the cutoff is still 
lacking.

\vskip1.cm
\noindent
{\bf \large Acknowledgments}
\vskip0.5cm
We are grateful to M. Bill\'o and M. Caselle for many enlightening
conversations.
\vskip0.5cm
\noindent
{\large \bf Appendix A}
\vskip0.5cm
In this appendix we calculate in the large $n$ limit some relevant quantities,
such as $\log d_r$ and $\frac{\ch_r(\mathbf{2})}{d_r}$, in a representation $r$
associated to a Young diagram whose rows and columns scale respectively as
$n^{\alpha}$ and $n^{1-\alpha}$.
This is not the most general case.
However we have already shown in Section 3 that in the
large $n$ limit  the most general Young diagram can be decomposed into a
discrete set of subdiagrams $Y_t$ (see Fig. \ref{Figure:young} and
related discussion), each 
scaling as above with a different power $\alpha_t$. The calculation that we are
going to present below will also apply to each $Y_t$, and the result for the
whole Young diagram is obtained by summing the contributions of the different
$Y_t$'s.
The shaded area of the Young diagram in \ref{Figure:young} gives
contributions which are 
subleading in the large $n$ and can be neglected.
Let us consider first a Young diagram where rows and columns scale respectively as
$n^{\alpha}$ and $n^{1-\alpha}$. It is convenient then to introduce the
following continuous variables:
\be
x = \frac{i}{n^{\alpha}},~~~~~~y = \frac{j}{n^{1-\alpha}}
\label{xy}
\eeq
and correspondingly
\be
f(x) = \frac{r_i}{n^{1-\alpha}},~~~~~~g(y) \equiv f^{-1}(y) = \frac{s_j}{n^{\alpha}}
\label{feg}
\eeq
where the derivatives $f'(x)$ and $g'(y)$ are everywhere negative or
null. The variable $x$ ranges from $0$ to a maximum value $x_{\rm
max}= f^{-1}(0)$ and similarly $y$ ranges from $0$ to $y_{\rm max}=f(0)$.
Then it is easy
to replace the discrete variables with the continuous ones in the expression
of $d_r$ and  $\frac{\ch_r(\mathbf{2})}{d_r}$ and obtain:
\be
\log d_r = \log n! - n \int_0^{ f^{-1}(0)} dx \int_0^{f(x)} dy
\log \left\{ n^{\alpha} \left( f^{-1}(y) - x \right) + n^{1-\alpha}
\left( f(x) - y \right) \right\}
\label{logdim}
\eeq
and
\be
\frac{\ch_r(\mathbf{2})}{d_r}= \frac{1}{1-1/n} \left[ n^{-\alpha}
\int_0^{ f^{-1}(0)} dx f(x)^2 -2 n^{\alpha-1} \int_0^{ f^{-1}(0)} dx
 x f(x) \right]
\label{car2}
\eeq
while the constraint  (\ref{boxesn}) in the large
$n$ limit becomes:
\be
\int_0^{ f^{-1}(0)} dx f(x) = \int_0^{f(0)} dy f^{-1}(y) = 1
\label{const}
\eeq
Keeping only the
terms of order $n \log n$ and $n$ in eq.s (\ref{logdim}) we have
for $\log d_r$, in the large $n$ limit:
\ba
\log d_r &= &\alpha  n \log n - n \left[ 1+ \int_0^{f^{-1}(0)} dx
f(x)\left( \log f(x) -1 \right) \right] ~~~~~~~\alpha<1/2 \nonumber \\
\log d_r &= &(1-\alpha)  n \log n - n \left[1+ \int_0^{f(0)} dy f^{-1}(y)
\left( \log f^{-1}(y) -1 \right) \right] ~~~~~~~\alpha>1/2
\label{logdim2}
\ea
Similarly we obtain for $\frac{\ch_r(\mathbf{2})}{d_r} $, keeping terms up to
order $1$:
\ba
\log \left[\frac{\ch_r(\mathbf{2})}{d_r} \right]&=&- \alpha \log n
+ \log \left[ \int_0^{ f^{-1}(0)}  dx f(x)^2 \right]~~~~~~~\alpha<1/2
 \nonumber \\  \log \left[\frac{\ch_r(\mathbf{2})}{d_r} \right]&=&(1-\alpha)
 \log n+ \log \left[ \int_0^{f(0)} dy f^{-1}(y)^2 \right]~~~~~~~\alpha>1/2
\label{car3}
\ea
Consider now the most general case, where the Young diagram consists, in the
large $n$ limit, of a discrete set of subdiagrams $Y_t$, each scaling with a
different power $\alpha_t$. Each $Y_t$ contributes to  $\log d_r$ with a term of
the form (\ref{logdim2}), with the appropriate $\alpha_t$ in place of $\alpha$
and weighted with its area $z_t$. The sum over $t$ reproduces, for the leading
$n \log n$ term, eq.(\ref{dir}).
Consider now $\frac{\ch_r(\mathbf{2})}{d_r}$.  It is clear from (\ref{car2})
that the asymptotic behavior of the contribution coming from $Y_t$ is
$n^{-\alpha_t}$ (resp. $n^{1-\alpha_t}$) for $\alpha_t<1/2$
(resp. $\alpha_t>1/2$).
The sum over $t$ is then dominated by subdiagrams with scaling powers $\alpha=0$
and $\alpha=1$ which are discussed in  detail in Section 3.

\vskip0.5cm
\noindent
{\large \bf Appendix B}
\vskip0.5cm
In this Appendix we derive Eqs.~(\ref{N1})  and (\ref{N1x}), in two
different ways: first by direct combinatorial methods, then
using the character expansion of the probability distribution discussed
in Subsec 4.1.
\vskip0.5cm
Consider a random walk in
$S_n$ in which, at each step, the permutation is multiplied by a
randomly chosen transposition with probability $x$, or by the identity
with probability $1-x$. It is convenient to think of $n$ objects and
$n$ boxes: initially the object $i$ is in the box $i$. Then we start
moving them around with the following rule: At each step, we exchange
two randomly chosen objects with probability $x$, or we do nothing
with probability $1-x$.
\par
Choose now one of the $n$ objects, say number 1, and compute the
probability $P_1(x,p)$ that, after $p$ steps, it is in box number 1 (either
because it never left it, or because it went back to it).
The expected number of cycles of length 1 in the final permutation is
then just
\be
N_1(x,p)=n P_1(x,p)
\eeq
To compute $P_1(x,p)$, write it as
\be
P_1(x,p)=\sum_{k=0}^p q(k) s(k)
\label{P1}
\eeq
where $q(k)$ is the probability that object 1 changes box exactly $k$
times during the walk, and $s(k)$ is the probability that it will be
back in its original box after changing box $k$ times.
We have easily
\be
q(k)=\left(p \atop k \right) \left(\frac{2x}{n}\right)^k
\left(1- \frac{2x}{n}\right)^{p-k}
\eeq
\par
For $s(k)$ we can write a recursion relation: suppose element 1 {\em
is} in box 1 after being moved $k$ times: Then it will certainly not
be in box 1 after being moved $k+1$ times. If it {\em is not} in box
$1$ after being moved $k$, times, it will be after being moved $k+1$
times with probability $1/(n-1)$. Hence
\be
s(k+1)=\frac{1}{n-1} \left[1-s(k)\right]
\eeq
that together with the initial condition $s(0)=1$ gives
\be
s(k)=\frac{1}{n}\left[ 1-\frac{1}{(1-n)^{k-1}}\right]
\label{s}
\eeq
Substituting in Eq.~(\ref{P1}) we obtain
\be
P_1(x,p)=\frac{1}{n}\left[1+(n-1)\left(\frac{n-2x-1}{n-1}\right)^p\right]
\eeq
(this result for $x=1-1/n$ was already quoted in
Ref.~\cite{Diaconis:1981}), and
\be
N_1(x,p)=1+(n-1)\left(\frac{n-2x-1}{n-1}\right)^p
\label{N1xx}
\eeq
which includes in particular Eq.~(\ref{N1}),
so that taking the limit $n\to\infty$ with $p=A n\log n$ one obtains
Eq.~(\ref{N1x}).
\vskip0.5cm
The same result can be obtained with character expansion methods by
noticing that $N_1(x,p)$ is simply the expectation value
$<Tr\ Q>$ of the trace of the permutation obtained after $p$ steps,
where the trace is taken in the ``fundamental''
$n$-dimensional representation: Such representation is reducible to a
direct sum of the trivial representation (Young diagram made of one
line of $n$ boxes) and the $n-1$ dimensional representation described
by a diagram with two rows of length $n-1$ and $1$. Therefore
\be
Tr\  Q=\ch_1(Q)+\ch_{n-1}(Q)
\eeq
\par
Using the character expansion of the probability distribution as in
subsec. 4.1 we obtain
\begin{eqnarray}
N_1(x,p)&=&\langle Tr\, Q \rangle\nonumber\\
&=&\frac{1}{n!}\sum_{Q\in
S_n}\left[\ch_1(Q)+\ch_{n-1}(Q)
\right]\sum_r d_r
\left[ (1-x) + x \frac{\ch_r(\mathbf{2})}{d_r} \right]^p
\ch_r(Q)\nonumber\\
&&
\end{eqnarray}
Using the orthogonality of characters and
\be
\frac{\ch_{n-1}(\mathbf{2})}{d_{n-1}}=\frac{n-3}{n-1}
\eeq
we find again Eq.~(\ref{N1xx}).

\end{document}